\documentclass[%
 reprint,
 superscriptaddress,
 amsmath,amssymb,
 aps,
]{revtex4-2}
\usepackage{graphicx}
\usepackage{dcolumn}
\usepackage{bm}
\usepackage{ulem}
\usepackage{soul}
\usepackage{color}
\usepackage{array}
\usepackage{booktabs}
\usepackage{multirow, makecell}
\usepackage{physics}
\usepackage{lineno}
\usepackage[authormarkup=none]{changes}
\usepackage[english]{babel}
\usepackage[colorlinks=true, allcolors=blue]{hyperref}%

\definechangesauthor[color=red]{BG}

\begin{document}

\preprint{APS/123-QED}

\title{Universal Signature of Hundness and Its Quantification}

\author{Dongwook Kim}\thanks{These authors contributed equally to this work.}
\affiliation{Department of Chemistry, Pohang University of Science and Technology, Pohang 37673, Republic of Korea}
\affiliation{Institute of Solid State Physics, TU Wien, 1040 Vienna, Austria}

\author{Ina Park}\thanks{These authors contributed equally to this work.}
\affiliation{Department of Chemistry, Pohang University of Science and Technology, Pohang 37673, Republic of Korea}
\affiliation{Center for Computational Quantum Physics, Flatiron Institute, New York 10010, USA}

\author{Bo Gyu Jang}
\email{bgjang@khu.ac.kr}
\affiliation{Department of Materials Science and Engineering, Kyung Hee University, Yongin 17104, Republic of Korea}

\author{Ji Hoon Shim}
\email{jhshim@postech.ac.kr} 
\affiliation{Department of Chemistry, Pohang University of Science and Technology, Pohang 37673, Republic of Korea}

\date{\today}

\begin{abstract}
Hund's coupling $J$ induces fundamentally different correlation effects from Hubbard $U$. This leads to a violation of the Brinkman--Rice scenario and the emergence of a Janus-faced phase owing to its low-energy effectiveness, in which band renormalization is confined below a characteristic energy scale. We propose a quantitative framework to capture low-energy effectiveness through two correlation factors: $z_L$ for low-energy quasiparticle renormalization and $z_H$ for high-energy charge fluctuations, newly introduced in this study. The discrepancy between $z_L$ and $z_H$ reflects the Hund character of the correlation. By establishing a one-to-one correspondence between correlation factors and the spin and charge susceptibilities, we identify the spin-degree-of-freedom effectiveness as the microscopic origin of low-energy effectiveness. Our framework, validated across multiorbital models and real materials, provides a universal and quantitative measure of Hundness.
\end{abstract}

\maketitle

\begin{figure*}
    \centering
    \includegraphics[width=0.9\linewidth]{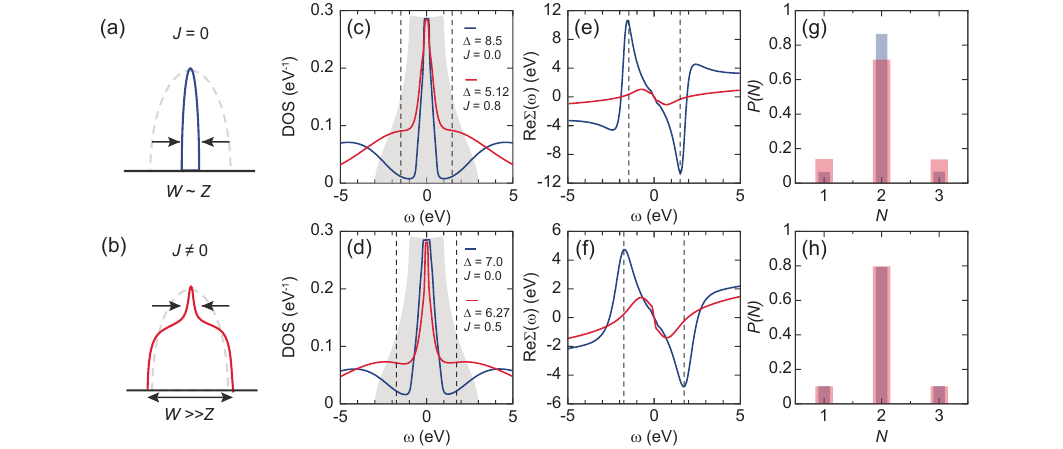}
    \caption{(a,b) Schematic quasiparticle density of states with (a) $J=0$ and (b) $J \neq 0$. $W$ is renormalized quasiparticle bandwidth. (c,d) Density of states, (e,f) real part of the self-energies, and (g,h) probability histogram of electron occupancies for the half-filled two-orbital systems. Upper panel is for the same renormalization factor $z_L = 0.2$, and lower panel is for the same \textcolor{black}{$P(2)$}. In (c,d), grey fills denote the bare DOS.}
    \label{fig1}
\end{figure*}

\textbf{\textit{Introduction}}--Metal–insulator transition (MIT) is one of the most fundamental phenomena in the physics of strong correlations.  A major success of dynamical mean-field theory (DMFT) was its ability to capture the essential features of the MIT~\cite{georges1992hubbard, georges1996dynamical}. As the Hubbard interaction $U$ increases, the overall conduction bandwidth is renormalized as Fig.~\ref{fig1}(a), and the spectral weight is transferred to the Hubbard bands.~\cite{zhang1993mott} This observation agrees with the widely accepted Brinkman–Rice (BR) picture~\cite{brinkman1970application}, which asserts the equivalence of quasiparticle (QP) and conduction band renormalization. Owing to this agreement, the BR scenario has long been regarded as a cornerstone \textcolor{black}{in the physics of strong correlation.}

However, this equivalence can break down in multiorbital systems~\cite{jang2021direct, park2024clean}. In the presence of Hund’s coupling $J$, only the narrow energy window near the Fermi level associated with QPs is strongly renormalized, while the overall conduction band undergoes weak renormalization, as illustrated in Fig.~\ref{fig1}(b). We refer to this behavior as the `low-energy effectiveness (LEE)' of Hund correlations\textcolor{black}{, where band renormalization is confined to a low-energy scale, leading} to the violation of the BR picture.
This indicates that Hund correlations and Mott-Hubbard correlations manifest in distinct manners~\cite{deng2019signatures, stadler2019hundness, stadler2021differentiating}.

Overall, Hund’s coupling leads to a characteristic separation between low- and high-energy correlations. This behavior, often discussed in the context of so-called Hund metals~\cite{haule2009coherence, stadler2021differentiating,yin2011kinetic, de2011janus, mravlje2011coherence, georges2013strong, stadler2015dynamical, stadler2019hundness,    wadati2014photoemission,horvat2016low,stricker2014optical,deng2016transport}, which often refers to multiorbital systems away from half-filling and single occupancy \textcolor{black}{such as iron-based superconductors, ruthenates, or molybdates~\cite{haule2009coherence, yin2011kinetic, mravlje2011coherence, wadati2014photoemission, deng2016transport, georges2013strong}}, demonstrates that the LEE of Hund correlations is not restricted to the half-filled case. In such regimes, one encounters a ``Janus-faced''~\cite{de2011janus} situation in which the QP renormalization factor $Z$ is strongly suppressed despite the system remains far from the Mott insulating state~\cite{de2011janus, georges2013strong, stadler2019hundness}. This disconnect highlights that $Z$, although being central parameter within the single-band BR scenario, is insufficient as a universal measure of correlated states or an indicator of Hund versus Mott-Hubbard correlation strength.

While this Hund-driven dichotomy has been qualitatively discussed in terms of spectral features~\cite{stadler2021differentiating, deng2019signatures, jang2021direct, park2024clean}, the understanding on it is still rudimentary due to the lack of quantitative and system-independent framework. In this study, we establish a universal framework by introducing a complementary high-energy correlation factor $z_H$ in addition to the conventional $Z$ (denoted as $z_L$ in this work). By clarifying \textcolor{black}{the microscopic origin of LEE} based on degenerate multiorbital Hubbard model and demonstrating its validity to the real materials, we provide a complete and universal scheme to quantify Hundness beyond the limitations of the BR picture.

\textcolor{black}{The main text focuses on DMFT results for the degenerate multi-orbital Hubbard model, primarily the half-filled two-orbital case. The same trend is also confirmed in the $1/3$-filled three-orbital case (2 electrons occupied in 3 orbitals), as shown in Secs.~3 and 4 of the SM~\cite{supp}, demonstrating that our discussion is not restricted to a specific filling. Furthermore, the applicability of this framework to real materials is demonstrated through a density functional theory (DFT) combined with DMFT. Details of the model and computational methods are provided in Sec. 1 of the SM~\cite{supp}.}

\textbf{\textit{Quantitative Framework}}--For the quantitative discussion on LEE, we introduce the two correlation factors, each representing the renormalization strength at low- and high-energy scale. For the low-energy scale, the QP renormalization factor \textcolor{black}{defined as
\begin{equation}
    \label{eq1}
    z_L \equiv Z = \frac{1}{m^*} = \bigg( 1 - \frac{\partial \mathrm{Re} \Sigma(\omega)}{\partial \omega}\bigg)^{-1}_{\omega = 0}
\end{equation}
by the frequency $\omega$ derivative of real part of the self-energy $\mathrm{Re}\Sigma(\omega)$,} naturally represents the renormalization strength. \textcolor{black}{Note that $m^*$ represents the QP mass enhancement.} Hereafter, we'll denote it as $z_L$, the low-energy correlation factor.

Comparing the two systems with equal $z_L$, the concept of LEE can be intuitively illustrated. Fig.~\ref{fig1}(c) shows the density of states (DOS) for the half-filled two-orbital systems with the same $z_L$ but different charge-transfer energy $\Delta$ \textcolor{black}{(see Sec. 2.A of the SM~\cite{supp} for details)} and $J$ where the red and blue lines indicate Hundless ($J/U = 0$) and Hund-correlated cases, respectively. At low energies, both systems exhibit identical peak widths near the Fermi level. However, at higher energies around 3 eV which corresponds to the half of the bare bandwidth $D$, a clear difference arises. As a result, the Hund-correlated system retains substantially more spectral weight within the bare bandwidth range, reflecting the manifestation of LEE in the context of bandwidth renormalization.

The LEE of bandwidth renormalization is fundamentally encoded in the real part of the self-energy, Re$\Sigma(\omega)$. As shown in Fig.~\ref{fig1}(e), the Hundless case exhibits an extended slope across the entire renormalized conduction band as marked by the dashed line. \textcolor{black}{Note that the dashed line corresponds to the frequency range where the Re$\Sigma(\omega)$ is nearly linear with a negative slope for Hundless case, indicating the renormalized conduction band range.} \textcolor{black}{Within this range, narrowing of the conduction band is nearly uniform, being consistent to the BR scenario.} In contrast, the Hund-correlated case exhibits a steep slope only within a narrow energy window, beyond which the slope flattens and the renormalization becomes weaker~\cite{wadati2014photoemission, jang2021direct, park2024clean, kugler2020strongly, stadler2021differentiating, chang2024dispersion}. 
This results in confined renormalization and ill-developed Hubbard bands \textcolor{black}{due to insufficient conduction band spectral weight loss.}

Since the development of Hubbard bands is associated with the suppression of charge fluctuation, \textcolor{black}{we compare the $P(N_{\mathrm{occ}})$, defined as the total probability of local atomic multiplets whose electron occupancy equals the nominal average occupation $N_{\mathrm{occ}}$ for these systems. (See Sec.~2.B of the SM~\cite{supp} for detailed definition on $P(N)$.) As shown in Fig.~\ref{fig1}(g), among the two systems with the same $N_{\mathrm{occ}} = 2$ and $z_L$, $P(2)$ is notably smaller in the Hund-correlated case, indicating larger charge fluctuation, consistent with the ill-developed Hubbard bands.}

\textcolor{black}{Next we compare the loss of conduction band spectral weight} for systems with the same \textcolor{black}{$P(2)$}. As shown in Fig.~\ref{fig1} (d), the remaining spectral weight in the renormalized conduction band range \textcolor{black}{for the Hundless case} marked by the dash line is nearly identical \textcolor{black}{for the systems with equal $P(2)$ shown in Fig.~\ref{fig1} (h),} establishing a clear correspondence between the loss of conduction band spectral weight and \textcolor{black}{$P(2)$}. On the other hand, the peak width at the Fermi level is narrower in the Hund-correlated case\textcolor{black}{, due to the smaller $z_L$ associated with the steeper Re$\Sigma(\omega)$ near the Fermi level as shown in Fig.~\ref{fig1}(f).}

\textcolor{black}{From the correspondence between $P(N_{\mathrm{occ}})$ and conduction band spectra,} we can make a quantitative discussion on LEE by comparing it to $z_L$. However, \textcolor{black}{$P(N_{\mathrm{occ}})$} alone cannot serve as a universal measure, as its lower bound depends on both the number of orbitals $M$ and the occupancy $N_{\mathrm{occ}}$. To construct a system-independent measure of high-energy correlation, we define a normalized $z_H$. In the atomic limit where the charge fluctuations are fully suppressed, the probability reaches its maximum value, $P_{atomic}(\textcolor{black}{N_{\mathrm{occ}}}) = 1$. In contrast, in the noninteracting \textcolor{black}{(free)} limit, all atomic multiplets are equally probable, yielding
\begin{equation}
\label{eq2}
\textcolor{black}{P_{\text{free}}(N_{\mathrm{occ}})} = \frac{{}^{2M}C_{N_{\mathrm{occ}}}}{2^{2M}}.
\end{equation}
In realistic systems it should lie between these two limits, with smaller  \textcolor{black}{$P(N_{\mathrm{occ}})$} indicating stronger charge fluctuations. The high-energy correlation factor $z_H$ is then defined through the linear interpolation between the atomic and non-interacting limits:
\begin{equation}
\label{eq4}
\begin{aligned}
P(\textcolor{black}{N_{\mathrm{occ}}}) &= z_H P_{\text{free}}\textcolor{black}{(N_{\mathrm{occ}})} + (1 - z_H) P_{\text{atomic}}\textcolor{black}{(N_{\mathrm{occ}})}, \\
\end{aligned}
\end{equation}
\textcolor{black}{so that $z_H$ represents the relative location of the system between the two limits. This yields}
\begin{equation}
\label{eq5}
\begin{aligned}
\textcolor{black}{z_H \equiv \frac{P(N_{\mathrm{occ}}) - P_{atomic}(N_{\mathrm{occ}})}{P_{\mathrm{free}}(N_{\mathrm{occ}}) - P_{atomic}(N_{\mathrm{occ}})} = \frac{1 - P(N_{\mathrm{occ}})}{1 - \frac{{}^{2M}C_{N_{\mathrm{occ}}}}{2^{2M}} } . }
\end{aligned}
\end{equation}

One of the key advantages of $z_H$ is that it is normalized in the same manner as $z_L$, with well-defined bounds in the range $z_H \in [0,1]$. Here, $z_H = 1$ corresponds to the least correlated limit, while $z_H = 0$ represents the most strongly correlated regime. However, there is a key distinction between the two factors, $z_L$ and $z_H$. While $z_L = 0$ implies an insulating state, $z_H = 0$ \textcolor{black}{implies} an isolated atom, and it remains finite even in insulating systems due to residual charge fluctuations.

\begin{figure}
    \centering
    \includegraphics[width=0.9\linewidth]{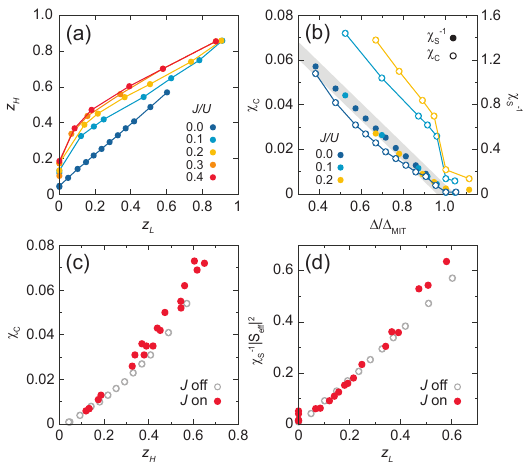}
    \caption{(a) $z_H$ vs $z_L$ and (b) $\chi_C$ ($\chi_S^{-1}$) vs $\Delta/\Delta_{\mathrm{MIT}}$ across the MIT with different $J/U$ ratios in half-filled two-orbital systems. $\Delta_{MIT}$ = 10.5, 6.555, 5.76 for $J/U$ = 0.0, 0.1 and 0.2 respectively. (c) $\chi_C$ vs $z_H$ \textcolor{black}{in the metallic regime} and (d) $\chi_S^{-1}|S_{\mathrm{eff}}|^2$ vs $z_L$ with and without $J$ in half-filled two-orbital systems.}
    \label{fig2}
\end{figure}

Figure~\ref{fig2}(a) shows the relation between $z_H$ and $z_L$ during the evolution toward the Mott insulating state for various values of $J/U$ ranging from 0 to 0.4 in half-filled two-orbital systems. In the Hundless case, the system follows the BR picture, exhibiting $z_H \approx z_L$ across the entire regime. In contrast, systems with finite $J$ exhibit a clear deviation: for the same value of $z_L$, systems with larger $J/U$ show higher values of $z_H$. This indicates that the renormalization becomes more confined to the low-energy sector \textcolor{black}{as the increase of Hund character, illustrating LEE quantitatively.}

These observations lead to two key insights. First, the BR picture can be reinterpreted as the limiting case where high- and low-energy renormalizations coincide, i.e., $z_H \approx z_L$. Second, the deviation between $z_H$ and $z_L$ provides a quantitative measure of the Hund character in multiorbital systems. \textcolor{black}{As shown in Fig.~S1 in the SM~\cite{supp}, these trends also hold in 1/3-filled three-orbital case, showing that the dichotomy between $z_H$ and $z_L$ also characterizes the Hund correlation in non-half-filled systems.}

\textbf{\textit{Microscopic Origin}}--A key advantage of the present definition of LEE is that it offers insight into its microscopic origin, providing a complementary perspective to the conventional spectral point of view. To elucidate the microscopic origin of LEE, we revisit the fundamental distinction between the roles of $U$ and $J$.

The Hubbard $U$ suppresses charge fluctuations by introducing an energy cost for electron hopping, which suppresses the Kondo exchange and in turn results in QP renormalization.
In contrast, Hund's coupling primarily suppresses intra-atomic fluctuations by lifting the ground-state degeneracy and locking electrons into high-spin states. This spin locking strongly renormalizes QP by reducing the coherence (Kondo) scale~\cite{nevidomskyy2009kondo,blandin1968magnetic, schrieffer1967kondo} while leaving the charge fluctuations largely unaffected. \textcolor{black}{In fact, Fig.~S5 in the SM~\cite{supp} evidences that $J$ is irrelevant to the charge fluctuations, showing that $z_H$ is almost constant with fixed $\Delta$. Fig.~S4 of the SM~\cite{supp} also shows that $z_H$ is $J$-independent throughout the correlated metallic regime under the evolution of $\Delta$.} In this sense, Hund's coupling has `spin-degree-of-freedom effectiveness (SDFE)' \textcolor{black}{in the metallic regime, being ineffective to the charge suppression while suppressing the spin fluctuation strongly, originating the} reduced coherence scale commonly observed in Hund's metals~\cite{haule2009coherence, yin2011kinetic, georges2013strong, de2011janus,  mravlje2011coherence, wadati2014photoemission, kang2021optical,  stadler2015dynamical, horvat2016low, deng2019signatures, stadler2019hundness, kugler2020strongly, stricker2014optical, deng2016transport}.

The SDFE of Hund's coupling can be clearly illustrated by the susceptibilities. Figure~\ref{fig2}(b) presents the evolution of charge and inverse spin susceptibilities, $\chi_C$ and $\chi_S^{-1}$, respectively, for fixed $J/U$ ratios ranging from 0.0 to 0.2. 
Across all the cases, $\chi_S^{-1}$ remains equivalent for a given $\Delta/\Delta_{\mathrm{MIT}}$ as indicated by the thick grey line.
In contrast, $\chi_C$ exhibits a clear dependence on $J/U$, with larger values of $J/U$ yielding higher $\chi_C$ at the same $\Delta/\Delta_{\mathrm{MIT}}$ reflecting the weaker suppression of the charge fluctuation. 

We next examine the quantitative relation between SDFE and LEE by directly comparing the susceptibilities to $z_H$ and $z_L$. Since $z_H$ naturally encodes charge fluctuations, it is expected to correlate with $\chi_C$. Indeed, Fig.~\ref{fig2}(c) shows a clear one-to-one correspondence between $\chi_C$ and $z_H$, independent of the $J/U$ ratio \textcolor{black}{in the metallic regime.} \textcolor{black}{Note that as the $\Delta$-linearity of $\chi_C$ shown in Fig.~\ref{fig2} (b), $z_H$ is also linear to $\Delta$ in the metallic regime due to the one-to-one correspondence between them, as shown in Fig.~S4 of SM~\cite{supp}.} 
\textcolor{black}{Likewise, $z_L$ corresponds to the strength of Kondo screening, and is thus expected to correlate with the inverse spin susceptibility normalized by the effective unscreened total spin square as $(\chi_{S}/|S_{\mathrm{eff}}|^2)^{-1}$. Here, $|S_{\mathrm{eff}}|^2 = 1/2$ for $J=0$ and $|S_{\mathrm{eff}}|^2 = 1$ for $J \neq 0$, based on the relevant multiplet average (see Sec.~2.C of the SM~\cite{supp} for details).}


The comparison between $\chi_{S}^{-1}|S_{\mathrm{eff}}|^{2}$ and $z_L$ is shown in Fig. 2 (d), where we find a clear one-to-one correspondence between $\chi_{S}^{-1}|S_{\mathrm{eff}}|^{2}$ and $z_L$. \textcolor{black}{This demonstrates that the QP renormalization $z_L$ is governed by spin suppression rather than charge suppression, in contrast to the expectation from the traditional BR picture.}
\textcolor{black}{Moreover, $\chi_C$-vs-$\chi_S^{-1}$ in Fig.~S2 of the SM~\cite{supp} closely resembles the $z_H$-vs-$z_L$ in Fig.~\ref{fig2}(a), explicitly demonstrating the SDFE for the half-filled case. A similar resemblance is also observed for the 1/3-filled three-orbital systems, between $z_H$-vs-$z_L$ (SM Fig.~S1)~\cite{supp} and $\chi_C$-vs-$\chi_S^{-1}$ (SM Fig.~S3)~\cite{supp}. These similarities, together with the one-to-one correspondences between the correlation factors and the susceptibilities, establish the connection between LEE and SDFE.} By bridging the spectral perspective on LEE with the susceptibility-based interpretation of SDFE through the \textcolor{black}{low- and high- energy} correlation factors, our framework thus provides the microscopic understanding on LEE, revealing the distinct roles of charge and spin correlations.

\begin{figure}
    \centering
    \includegraphics[width=0.9\linewidth]{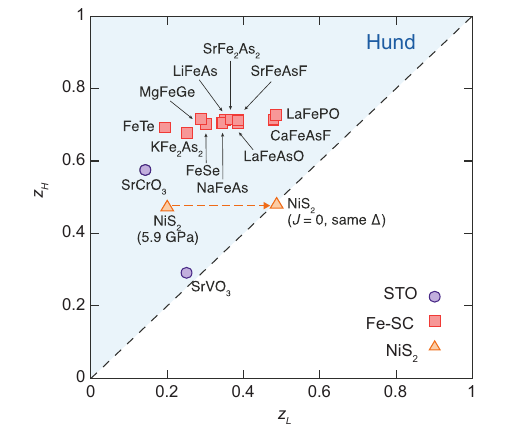}
    \caption{Real materials in $z_H$ vs $z_L$ phase diagram based on DFT+DMFT results with various number of orbitals and electron occupancies. Triangles are iron-based high-$T_\text{c}$ superconductors (Fe-SC), squares are NiS$_2$ under $P$ = 5.9 GPa with different interaction parameters, and circles are strontium transition-metal (TM) oxides (STO) with chemical formula of Sr(TM)O$_3$. The dashed line marks the $z_H=z_L$ line.}
    \label{fig3}
\end{figure}

\textbf{\textit{Application to Real Materials}}--Finally, we demonstrate the applicability of $z_H$-vs-$z_L$ framework as a quantitative measure of Hund character in real materials, thereby providing strong support for its relevance beyond model systems. Figure~\ref{fig3} summarizes the result obtained from DFT+DMFT for the representative multiorbital compounds that have been extensively investigated in the context of Hund-driven correlations, including the iron-based high-$T_\text{c}$ superconductors \cite{haule2009coherence, yin2011kinetic, lanata2013orbital, schafgans2012electronic}, strontium transition metal oxides \cite{mravlje2011coherence,wadati2014photoemission,dang2015electronic,deng2019signatures}, and NiS$_2$.

Unlike the \textcolor{black}{degenerate multi-orbital model discussed above,} the $d$-orbitals (or a subset of the $d$-orbitals near the chemical potential) can be non-degenerate \textcolor{black}{in real materials}, giving multiple values of $z_L$. \textcolor{black}{To represent the low-energy correlation of the system,} we define the weighted average value $z_L^{avg}$ for all five $d$-orbitals of given transition metal atom as
\begin{equation}
    z_L^{avg} = \frac{\sum_i w_i z_L^i}{\sum_i w_i}
    \label{eq_zavg}
\end{equation}
where $z_L^i$ and $w_i$ represent the renormalization factor and the spectral weight at the Fermi level of $i$-th orbital, \textcolor{black}{each obtained from Eq.~\ref{eq1} and} 
\begin{equation}
    w_i \equiv N_i(E_\mathrm{F}) = -\frac{1}{\pi} \textrm{Im} [G\textcolor{black}{_{i}}(\omega = 0 )]. 
\end{equation} 
\textcolor{black}{where $G_{i}(\omega )$ represents the local Green's function of $i$-th orbital.} The self-energy and local Green's function were collected from the DMFT-LMTO Online Repository in Rutgers DFT\&DMFT Materials Database~\cite{RutgersDMFT} for iron-based superconductors and SrVO$_3$ and from Ref.~\cite{park2024clean} for NiS$_2$ under pressure. Those for SrCrO$_3$ were newly calculated in this study. \textcolor{black}{See Sec. 1.B of SM~\cite{supp} for details.}

\textcolor{black}{In the resulting phase diagram,} it is evident that all the materials investigated in this study, except for SrVO$_3$ and NiS$_2$ without $J$, are located in the upper left triangle of the phase diagram, namely the Hund region. This region indicates the exhibition of LEE, \textcolor{black}{i.e., $z_H > z_L$.} \textcolor{black}{In case of SrVO$_3$, it shows $z_H \approx z_L$ as the ground state is irrelevant to Hund exchange since the electron occupancy $N_{\mathrm{occ}} = 1$.} This observation is consistent with previous model studies~\cite{de2011janus}, which highlights the contrasting behaviors of SrVO$_3$ and SrCrO$_3$ (two electrons in $t_{2g}$ shell, corresponding to 1/3-filled three-orbital), where the latter clearly manifests the `Janus-faced'~\cite{de2011janus, georges2013strong} nature of Hund's coupling.

On the other hand, \textcolor{black}{in the material with $N_{\mathrm{occ}} \neq 1$,} we can turn off the $J$ in the DFT+DMFT calculation to exclude the Hund effect. The comparison between the two NiS$_2$ calculations with different $J$ values illustrates how the system evolves when $J$ is turned off while keeping the same atomic charge transfer energy, $\Delta = 9$ eV. (see orange dashed arrow) \textcolor{black}{The two NiS$_2$ results show the same $z_H$ while $z_L$ being strongly suppressed only with finite $J$. This is also consistently observed in BaOsO$_3$, a Hund metal with strong SOC~\cite{bramberger2021baoso}, as shown in Sec.~6 of the SM~\cite{supp}. Taken together, the DFT+DMFT results for prototypical correlated materials illustrate the LEE as a manifestation of Hund correlation, i.e., $z_H > z_L$.}


\textcolor{black}{Important remaining questions include the identification of experimental fingerprints of Hund correlations. While $z_L^{avg}$, associated with the average quasiparticle mass enhancement, can be measured from specific heat, $z_H$ reflects the local charge-state distribution $P(N_{\mathrm{occ}}\pm1)$, which is accessible to core-level spectroscopies such as X-ray photoemission/absorption spectroscopy (XPS/XAS). Furthermore, extending this framework to high-$T$ or non-equilibrium conditions is a challenging yet important direction for future study. For example, at high $T$, quasiparticles become ill-defined and strong low-energy correlations typically manifest as a small coherence temperature $T_K$. In this regime, the decay rate of the dynamical impurity susceptibility could potentially serve as a proxy for the low-energy correlation strength, given that it inherently encodes $T_K$.}

In this paper, we established a quantitative framework for understanding the low-energy effectiveness (LEE) of Hund's interaction by introducing two complementary correlation factors: the conventional low-energy correlation factor $z_L$ and a newly defined high-energy correlation factor $z_H$. From this framework, we could rephrase the LEE as a smaller $z_L$ relative to $z_H$. This framework also allows us to understand the microscopic origin of LEE, which has been discussed based on the macroscopic phenomena, i.e. violation of the Brinkman-Rice scenario. Introducing the concept of spin-degree-of-freedom effectiveness (SDFE) and showing its correspondence to the LEE, we could explain why the LEE of Hund correlation should be universal. Finally, the validity of this framework has been confirmed for real materials, where strongly Hund-correlated systems exhibiting pronounced $z_H$-$z_L$ discrepancy, supporting the universal applicability of our approach.

\begin{acknowledgments}
We thank A. Millis, H. LaBollita\textcolor{black}{, M. Capone, and S. Giuli} for fruitful discussions. The Flatiron Institute is a division of the Simons Foundation.
\end{acknowledgments}

\nocite{*}

\bibliographystyle{apsrev4-2}  
\bibliography{references}      

\clearpage                
\thispagestyle{empty}     
\cleardoublepage          

\onecolumngrid

\renewcommand{\thefigure}{S\arabic{figure}}
\renewcommand{\theequation}{S\arabic{equation}}
\renewcommand{\thesection}{\arabic{section}}
\setcounter{figure}{0}
\setcounter{equation}{0}
\setcounter{section}{0}

\begin{center}
    \vbox{\vspace*{1cm}}
    {\large\bfseries Supplementary Material} \\[0.3cm]
    {\Large\bfseries \textit{``Universal Signature of Hundness and Its Quantification"}} \\[0.5cm]
    \vspace*{1cm}
\end{center}

\clearpage

\section{Details on Model and Methods}

\subsection{Model DMFT Calculation}

For the model calculation, degenerate multiorbital Hubbard model in the cubic lattice with only the nearest-neighbor hopping of $t = 0.5$ eV was used. For the presented results, temperature was fixed to $k_BT=1/125$ which was sufficiently lower than the coherence temperature in the metallic regime. On orbital configuration, half-filled two-orbtal systems and 1/3-filled three orbital systems were considered, both having the electron occupancy of 2.

The impurity problems are solved using the continuous-time quantum Monte Carlo method~\cite{gull2011continuous, werner2006continuous, haule2007quantum}, where the local Coulomb interaction Hamiltonian is written as
$$
H_{\rm int}^{(N_{\rm orb})}
=
H_{\rm dens}
+
H_{\rm sf}
+
H_{\rm ph},
$$
where $N_{\rm orb}$ denotes the number of the orbitals and the density-density part $H_{\rm dens}$ is
$$
H_{\rm dens}
=
U_p \sum_m n_{m\uparrow} n_{m\downarrow}
+
U'_p \sum_{m<m'} \sum_{\sigma}
n_{m\sigma} n_{m',-\sigma}
+
\left(U'_p-J^{\rm dd}_p\right)
\sum_{m<m'} \sum_{\sigma}
n_{m\sigma} n_{m'\sigma},
$$
the spin-flip $H_{\rm sf}$ part is
$$
H_{\rm sf}
=
-
J^{\rm sf}_p
\sum_{m\ne m'}
c^\dagger_{m\uparrow}
c_{m\downarrow}
c^\dagger_{m'\downarrow}
c_{m'\uparrow},
$$
and the pair-hopping part $H_{\rm ph}$ is
$$
H_{\rm ph}
=
J^{\rm ph}_p
\sum_{m\ne m'}
c^\dagger_{m\uparrow}
c^\dagger_{m\downarrow}
c_{m'\downarrow}
c_{m'\uparrow}
$$
and the coefficients $U_p$, $U'_p$, $J^{\rm dd}_p$, $J^{\rm sf}_p$, and $J^{\rm ph}_p$ are parametrized in $p$-orbital parametrization as
$$
U_p = U+\frac{4}{5}J,
\qquad
U'_p = U-\frac{2}{5}J,
\qquad
J_p^{\rm dd}=J_p^{\rm sf}=J_p^{\rm ph}=\frac{3}{5}J.
$$

For the three-orbital calculations, we use the full three-orbital problem with $N_{\rm orb}=3$. For the two-orbital calculations, we use $N_{\rm orb}=2$, obtained by removing one orbital from the same local interaction parametrization. Thus, the two-orbital model is not an independently parametrized two-orbital Kanamori Hamiltonian, but rather a reduced two-orbital version of the same Hund-coupled local problem used in the three-orbital case.

Note that since the $p$-shell parametrization and the conventional $t_{2g}$ Kanamori parametrization have different $J$-dependent coefficients, calculations performed with the same nominal values of $U$ and $J$ yield different results.

Although the local interaction is parametrized using the rotationally
invariant $p$-shell form, the parameter regime considered here is
motivated by correlated $d$-electron systems. In particular, the sizes
of $U$ and $J$ are chosen to represent a Hund-coupled multi-orbital
system with correlated $d$-orbital, rather than to model an
actual atomic $p$ shell. The purpose of the present calculations is
not to figure out how the physical quantities will be at specific $J$, but to provide a controlled benchmark for the physics of correlated $d$-orbital shell, where the cases of $N_{\rm orb}=2$ and $N_{\rm orb}=3$ correspond to the $e_g$-correlated and $t_{2g}$-correlated systems due to the crystal field splitting respectively.

\subsection{Real Materials DFT+DMFT Calculation}

For real materials data shown in this work, full charge-self-consistent DFT+DMFT calculation was done. First, DFT calculation was done using the \texttt{WIEN2k}, which employs the full-potential augmented plane wave method~\cite{Blaha2020}. On top of an effective one-electron Hamiltonian obtained from the \texttt{WIEN2k} calculation, charge self-consistent DFT+DMFT calculations were performed as implemented in the DFT + embedded DMFT functional (eDMFTF) code~\cite{haule2010dynamical, haule2015free}. For the DMFT calculations, a real harmonics basis for full five $d$ orbitals was used. (For the cases with spin-orbit coupling, such as BaOsO$_3$, we used $JJ$ basis.) In cases with local ligand field environment, further local basis transformation which maximally diagonalizes the local Hamiltonian was applied.

For the interaction Hamiltonian, we used the density-density form of the Coulomb interaction with Slater parametrization, $F^0 \equiv U$, $F^2 \equiv \frac{112}{13}J$, and $F^4 \equiv \frac{70}{13}J$ with interaction parameters $U$ and $J$ as tabulated in Table~\ref{tab:uj}. For materials Fe-SC family and SrVO$_3$, we used the same interaction parameters reported in the DMFT-LMTO online repository in the Rutgers DFT\&DMFT Materials Database~\cite{RutgersDMFT}, which were also used in the systematic comparison between Fe-SC materials in Ref~\cite{yin2011kinetic}. The $U$ and $J$ parameter used for NiS$_2$ were benchmarked from Ref~\cite{park2024clean}, which reproduced the phase diagram of NiS$_2$ qualitatively well.
The hybridization window was set from -10 eV to 10 eV, and the impurity problem was solved using a continuous-time quantum Monte Carlo (CTQMC) impurity solver. For calculations with spin-orbit coupling, we used a full interaction Hamiltonian which also contains spin-flip and pair-hopping interaction terms.

\begin{table}
\caption{\label{tab:uj}$U$ and $J$ interaction parameters used in DFT+DMFT calculations.}
\begin{ruledtabular}
\begin{tabular}{cccc | cccc}
Material & $U$ (eV) & $J$ (eV) & Reference & Material & $U$ (eV) & $J$ (eV) & Reference \\
\hline
NiS$_2$ & 8.0 & 1.0 & ~\cite{park2024clean} & & & & \\
\hline
SrCrO$_3$ & 8.0 & 0.7 & $^{\rm{a}}$ & SrVO$_3$ & 8.0 & 0.7 & ~\cite{RutgersDMFT} \\
\hline
LaFePO & 5.0 & 0.8 & ~\cite{RutgersDMFT, yin2011kinetic} & LaFeAsO & 5.0 & 0.8 & ~\cite{RutgersDMFT, yin2011kinetic} \\
CaFeAsF & 5.0 & 0.8 & ~\cite{RutgersDMFT, yin2011kinetic} & SrFeAsF & 5.0 & 0.8 & ~\cite{RutgersDMFT, yin2011kinetic} \\
NaFeAs & 5.0 & 0.8 & ~\cite{RutgersDMFT, yin2011kinetic} & LiFeAs & 5.0 & 0.8 & ~\cite{RutgersDMFT, yin2011kinetic} \\
SrFe$_2$As$_2$ & 5.0 & 0.8 & ~\cite{RutgersDMFT, yin2011kinetic} & KFe$_2$As$_2$ & 5.0 & 0.8 & ~\cite{RutgersDMFT, yin2011kinetic} \\
FeTe & 5.0 & 0.8 & ~\cite{RutgersDMFT, yin2011kinetic} & FeSe & 5.0 & 0.8 & ~\cite{RutgersDMFT, yin2011kinetic} \\
MgFeGe & 5.0 & 0.8 & ~\cite{RutgersDMFT, yin2011kinetic} &  & & &  \\
\hline
BaOsO$_3$ & 4.5 & 0.8 & &  & & & 
\footnotetext{We used the same $U$ and $J$ parameters as SrVO$_3$ for systematic comparison.}
\end{tabular}
\end{ruledtabular}
\end{table}

\clearpage

\section{Discussions based on the atomic multiplets in the Fock basis}

\subsection{Definition of interatomic charge-transfer energy $\Delta$}
The interatomic charge-transfer energy in the atomic limit is the energy required to move one electron from one to the other site, thus is defined as
\[
\Delta \equiv E(N_{\mathrm{occ}}+1)+E(N_{\mathrm{occ}}-1)-2E(N_{\mathrm{occ}}),
\]
where $E(N)$ is the minimum energy of the multiplets with occupancy $N$, and $N_{\mathrm{occ}}$ is the average occupancy of the system. In case of the half-filled two-orbital systems that this paper mainly discuss about, $\Delta = U +1.4J$ and in case of  2/3-filled three-orbital systems that we show the result in the supplement, $\Delta = U - J$. Due to the difference of signs in front of $J$, analysis based on the $(U, J)$ leads to the occupancy-dependent opposite influence on charge suppression from $J$. For the discussion on general occupancy-independent signature of Hund correlation, we take the central parameter set as $(\Delta, J)$, so that changing $J$ with fixed $\Delta$ doesn't change the charge suppression (Hubbard correlation), i.e., $J$ being disentangled to the charge suppression.

\subsection{Definition of $P(N)$}

Atomic multiplets are represented by the local Fock states $|N,\alpha\rangle$, where $N$ denotes the local occupation number and $\alpha$ labels the remaining quantum numbers. From the Fock basis, the ground state is expanded as $|\mathrm{gs}\rangle = \sum_{N,\alpha} A_{N\alpha} |N,\alpha\rangle$. Then $P(N)$, the probability of finding the atomic multiplets with occupation number of $N$ in the ground state can be obtained by summing $|A_{N\alpha}|^2$ over $\alpha$, i.e., $P(N) = \sum_{\alpha} |A_{N\alpha}|^2$. Note that $\sum_{N} P(N) =  \sum_{N,\alpha} |A_{N\alpha}|^2 = 1$ by normalization.

\subsection{Effective unscreened total spin square $|S_{\mathrm{eff}}|^2$}
The magnetic response of an isolated spin (local spin susceptibility) of $S$ follows the Curie's law, yielding $|S|^2$-proportional spin susceptibility. Thus, we can estimate the development of Kondo singlet by dividing local spin susceptibility with $|S|^2$.

In an impurity, or local site described as an impurity embedded in self-consistent bath, effective unscreened total spin square $|S_{\mathrm{eff}}|^2$ can be obtained by the expectation value of the $\hat{\mathbf{S}^2} = \hat{S_z}\hat{S_z} + \hat{S_+}\hat{S_-} + \hat{S_-}\hat{S_+}$ among the atomic multiplets in $N=N_{\mathrm{occ}}$ manifolds. For the Hundless case, since all the Fock states in $N=N_{\mathrm{occ}} =2$ manifold are equivalently favorable, $|S_{\mathrm{eff}}|^2 = 1/2$. For the Hund correlated cases, only the triplet states are favored since the temperature scale of $k_BT=1/125$ eV was order-of-magnitude smaller than the scale of $J$. Thus, $|S_{\mathrm{eff}}|^2 = 1$ if $J \neq 0$.

\clearpage

\section{Validity of $z_H$ and $z_L$ based framework on low-energy effectiveness in 1/3-filled three-orbital system.}

\begin{figure}[h]
    \centering
    \includegraphics[width=0.6\linewidth]{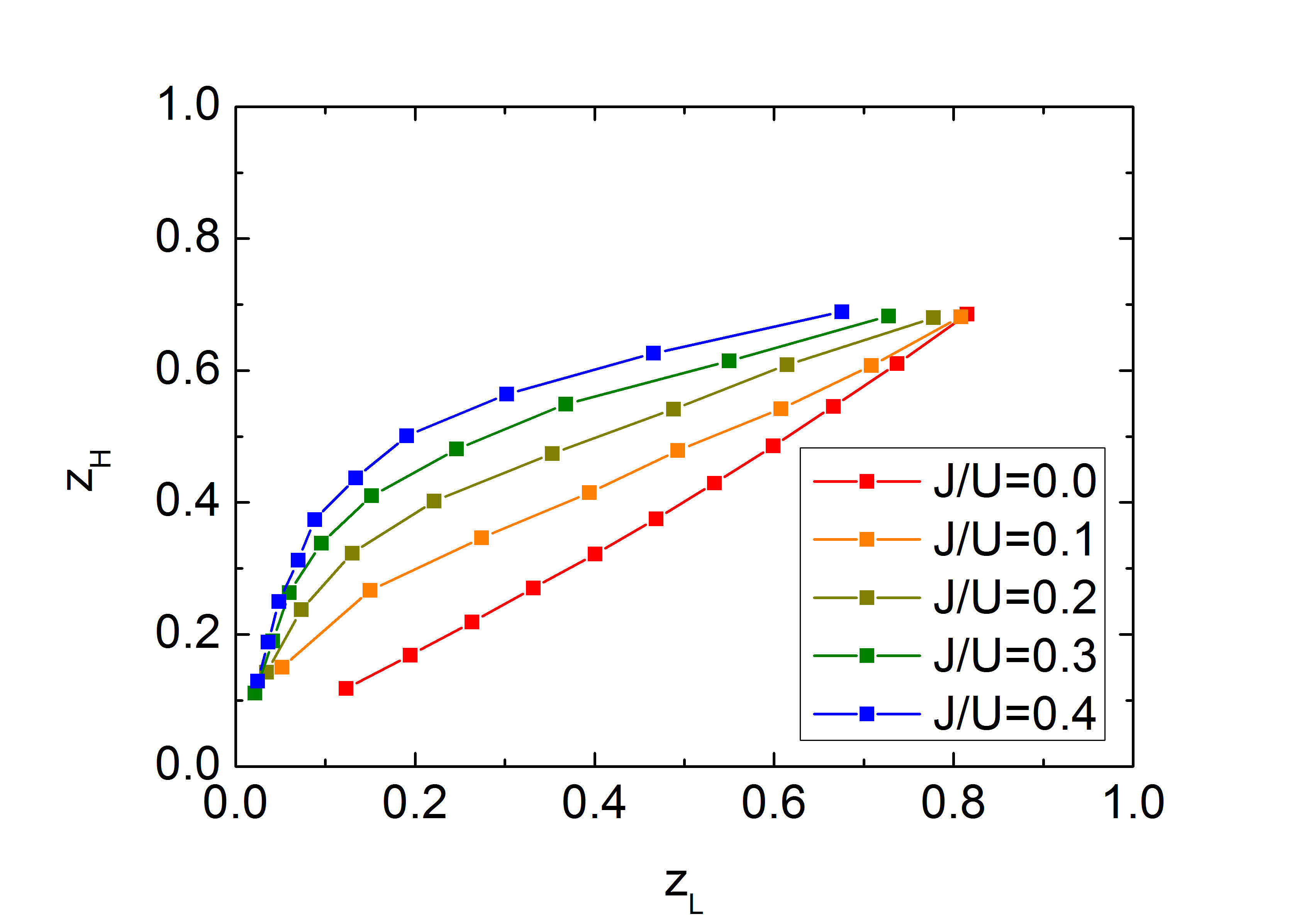}
    \caption{$z_H$ vs $z_L$ in 1/3-filled three-orbital system. As in the case of half-filled system, deviation between $z_H$ and $z_L$ increases as the increase of $J/U$. At $J = 0$ it follows the Brinkman-Rice picture, $z_H \approx z_L$.}
    \label{figS1}
\end{figure}

Similar to the case for the half-filled two-orbital system (2 electrons in 2 orbitals) shown in Fig.~2 (a), $z_H$ vs $z_L$ for 1/3-filled three-orbital system (2 electrons in 3 orbitals) in Fig. S1 shows similar behavior. In case where $J = 0$, the evolution as the increase of $U$ follows the $z_H \approx z_L$ line, consistent to the Brinkmann-Rice picture. However, as the increase of $J/U$, it moves away from the $z_H \approx z_L$. Therefore among the systems with equal $z_L$, $z_H$ gets larger as the $J/U$, the Hund character increases, reflecting the low-energy effectiveness (LEE). From this we show that the dichotomy between $z_H$ and $z_L$ as the increase of $J$ is also valid in non-half-filled case.

\clearpage

\section{Illustration of spin-degree-of-freedom effectiveness by $\chi_C$ vs $\chi_S^{-1}$}

\begin{figure}[h]
    \centering
    \includegraphics[width=0.6\linewidth]{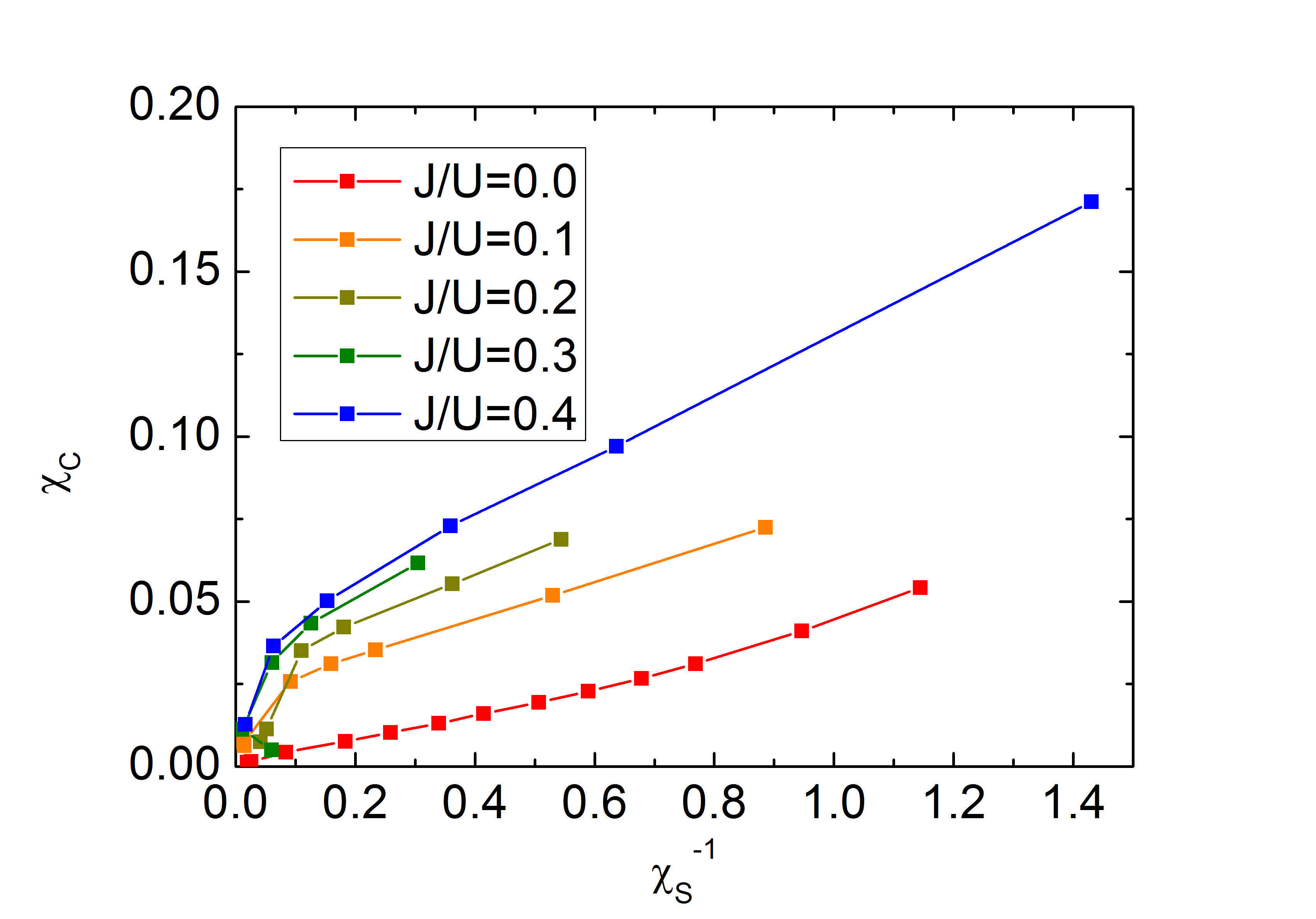}
    \caption{$\chi_C$ vs $\chi_S^{-1}$ in half-filled two-orbital system. Similar to the case of $z_H$ vs $z_L$, systems with $J=0$ shows simultaneous decrease as the evolution to metal-insulator transition. In case of finite $J$, it shows significant $\chi_C$ in comparison to decreased $\chi_S^{-1}$. As the increase of $J/U$ (Hund character) the discrepancy between $\chi_C$ and $\chi_S^{-1}$ increases.}
    \label{figS2}
\end{figure}

\begin{figure}[h]
    \centering
    \includegraphics[width=0.6\linewidth]{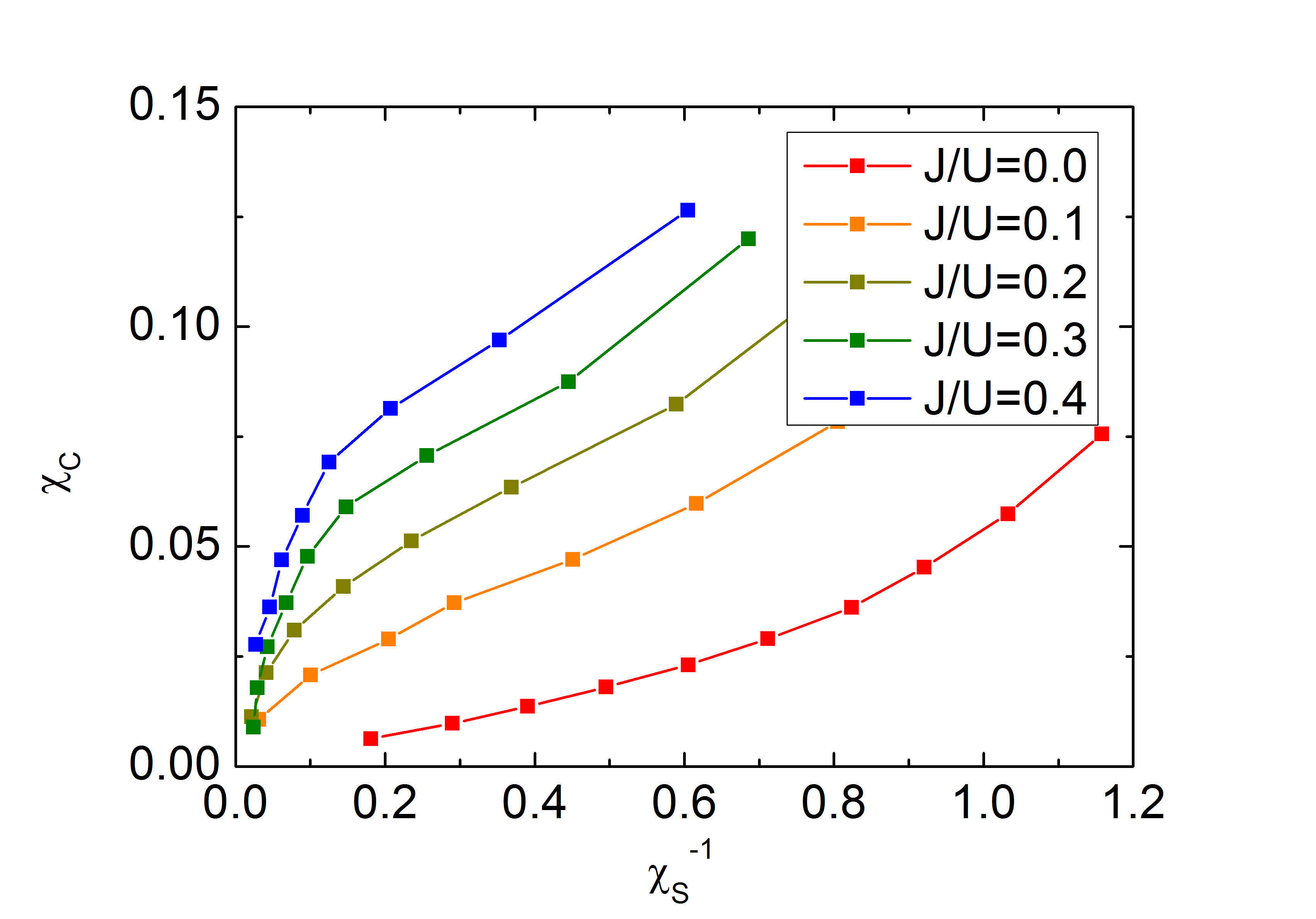}
    \caption{$\chi_C$ vs $\chi_S^{-1}$ in 1/3-filled three-orbital system. As in the case of a half-filled case, it shows similarity to $z_H$ vs $z_L$ such that the increase of $J/U$ (Hund character) the discrepancy between $\chi_C$ and $\chi_S^{-1}$ increases.}
    \label{figS3}
\end{figure}

$\chi_C$ vs $\chi_S^{-1}$ is shown in Fig. S2 and Fig. S3, for the half-filled two-orbital system and 1/3-filled three-orbital system respectively. Among the systems with the same $\chi_S^{-1}$, $\chi_C$ increases as the increase of the $J/U$ (Hund character), illustrating the spin-degree-of-freedom effectiveness (SDFE) that the Hund correlation strongly suppresses the spin fluctuation while not suppressing the spin fluctuation. The $\chi_C$ vs $\chi_S^{-1}$ resembles the $z_H$ vs $z_L$ in Fig. 2 (a) and Fig. S1 for half-filled two-orbital case and 1/3-filled three-orbital case respectively, visualizing the correspondence between LEE and SDFE.

\clearpage

\section{$\Delta$-linearity and $J$-constancy of $z_H$}

\begin{figure}[h]
    \centering
    \includegraphics[width=0.6\linewidth]{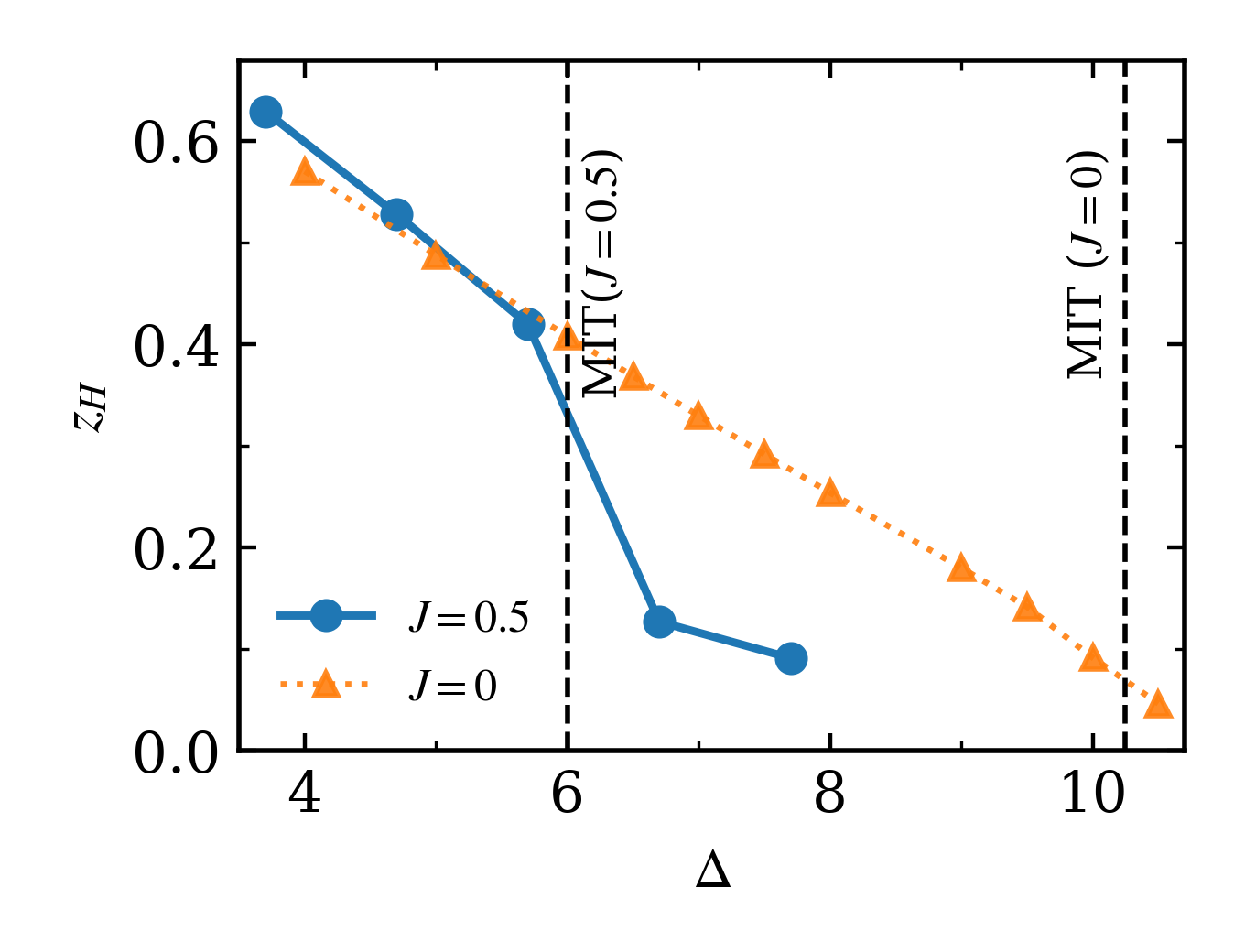}
    \caption{The evolution of $z_H$ as $\Delta$ with fixed $J$ of 0 and 0.5 for the half-filled two-orbital system. $z_H$ was nearly equal with the equal $\Delta$ for the different $J$ in the metallic phase.}
    \label{figS4}
\end{figure}

\begin{figure}[h]
    \centering
    \includegraphics[width=0.6\linewidth]{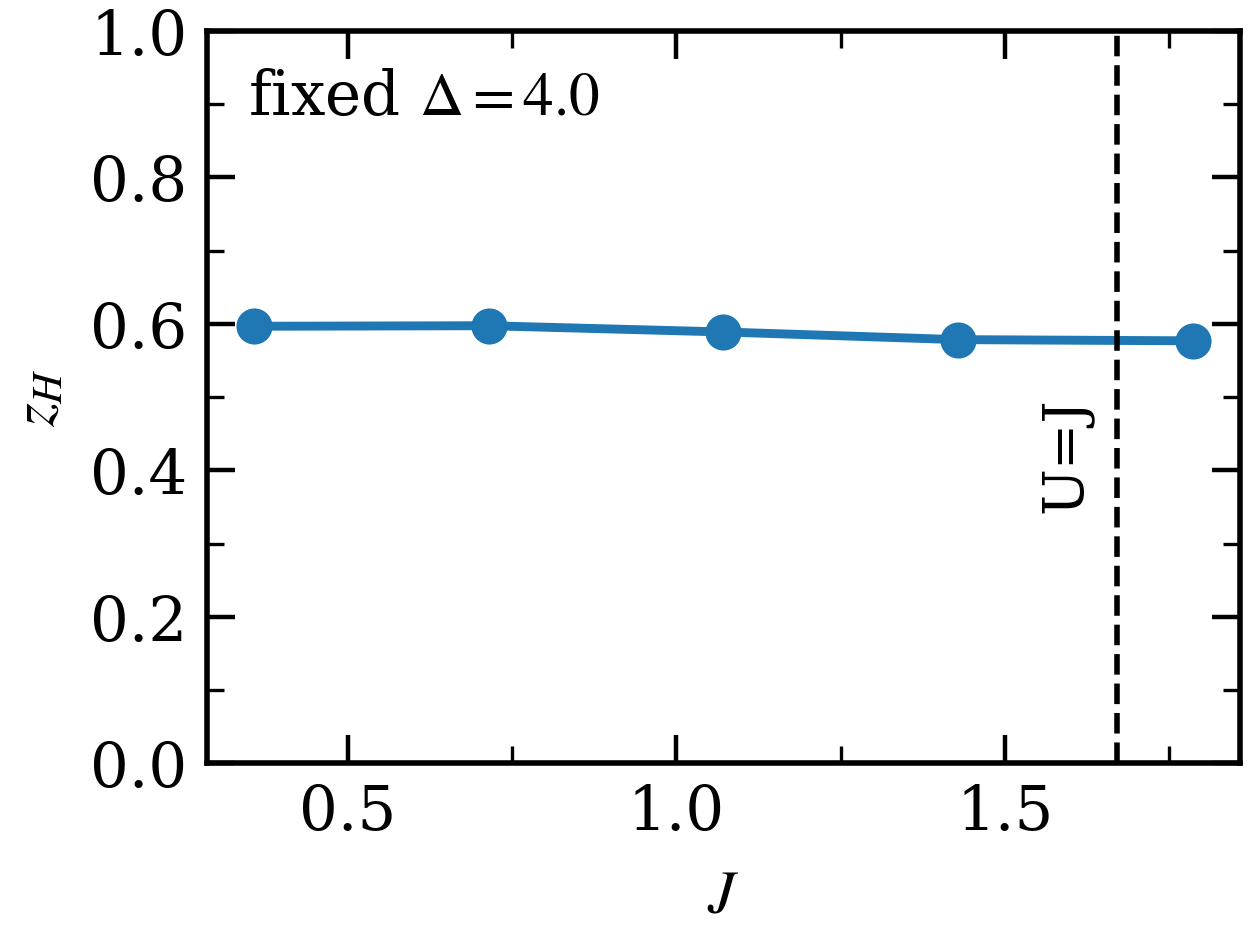}
    \caption{The evolution of $z_H$ as $J$ with fixed $\Delta=4.0$ for the half-filled two-orbital system. Dashed vertical line denote the $J$ value that becomes equal to $U$, beyond which is unrealistic regime. $z_H$ is nearly constant. Note that no MIT occurs in this $\Delta$.}
    \label{figS5}
\end{figure}

For the evidence that $J$ is nearly not affecting to the high-energy correlation and spin suppression, we show the evolution of $z_H$ as $\Delta$ with fixed $J$ of 0 and 0.5 in the half-filled two-orbital system in Fig. S4. In both cases, $z_H$ was equal for the equal $\Delta$ in the metallic regime. Thus the LEE and SDFE is strictly valid only for the metallic phase.

To clarify the $J$-independence of $z_H$, we show the evolution of $z_H$ as $J$ with fixed $\Delta=4.0$ in the half-filled two-orbital system in Fig. S5. Indeed, as metal-insulator transition (MIT) didn't occur in this case, $z_H$ shows no clear $J$-dependence in the metallic phase. Note that the regime above the $J$ marked as $U = J$ is unrealistic regime.

\clearpage

\section{Hundness with strong spin-orbit coupling}

\begin{figure}[h]
    \centering
    \includegraphics[width=0.5\linewidth]{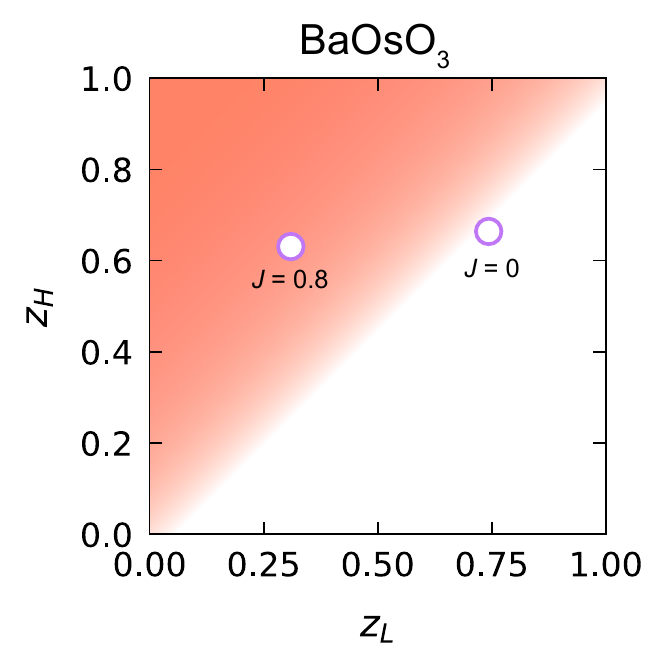}
    \caption{$z_H$ vs $z_L$ phase diagram for BaOsO$_3$. DFT+DMFT calculation was done including spin-orbit coupling, i.e. using $J_{\mathrm{tot}}=1/2, 3/2$ basis, with Hund's coupling $J$ turned on (0.8 eV) and off. $U$ was used as 4.5 eV.}
    \label{figS6}
\end{figure}



In case where strong spin-orbit coupling (SOC) exists, electrons are in the $JJ$ (total moment) basis instead of $LS$ (spin and orbital) basis since the total moment $J_{\mathrm{tot}}=L+S$, where $L$ and $S$ denotes the orbital and spin moment respectively, serves as a good quantum number. (For the consistency, we keep $J$ as referring to Hund's exchange and represent the total moment as $J_{\mathrm{tot}}$) In such case, the $z_L$ should rather be compared to the $\chi_{J}$, the susceptibility for $J_{\mathrm{tot}}$ than the spin susceptibility. Thus the spin-degree-of-freedom effectiveness should be modified based on $J_{tot}$, and whether the 1-to-1 correspondence between the $\chi_{J}$ and $z_L$ will be kept needs further verification.

However, underlying mechanism of Hundness from preference on higher momentum states is still valid. Consindering Osmates or Irridates which has $5d^4$ or $5d^5$, without the Hund $J$, $J_{\mathrm{tot}}$ singlet is preferred among the $J_{\mathrm{tot}}=3/2$ manifold. As Hund $J$ is included, it breaks the $J_{\mathrm{tot}}$ singlet and increases the atomic total moment.

In case of $\mathrm{BaOsO_3}$, OS-$5d^4$ configuration partially occupies both $J_{\mathrm{tot}}=1/2$ and $J_{\mathrm{tot}}=3/2$ due to relatively larger bandwidth compared to the SOC splitting. Thus it is a metallic system with bands corresponding to both $J_{\mathrm{tot}}=1/2$ and $J_{\mathrm{tot}}=3/2$ residing near the Fermi level. In such case, Hund $J$ leads to the increase of $J_{\mathrm{tot}}$ by moving the electron from $J_{\mathrm{tot}}=3/2$ to $J_{\mathrm{tot}}=1/2$ manifold~\cite{bramberger2021baoso}. DFT+DMFT results for the $\mathrm{BaOsO_3}$ done in the $JJ$ basis, shown in Fig.~S6, show the strong decrease of $z^{avg}_L$ with $z_H$ being nearly constant as the increase of $J$, demonstrating the validity of our framework under the existence of the strong SOC.

If the SOC splitting is larger than the bandwidth, bands corresponding to the $J_{\mathrm{tot}}=1/2$ and $J_{\mathrm{tot}} = 3/2$ manifold gets seperated. In this case, $5d^4$ configuration leads to the band insulator with only $J_{\mathrm{tot}}=3/2$ being fully occupied and $5d^5$ configuration leads to the effectively half-filled single-orbital Mott-Hubbard type system for $J_{\mathrm{tot}}=1/2$ orbital. Irridate materials correspond to these cases, which don't belong to the validity range of our framework.

\end{document}